\documentstyle[twoside,fleqn,espcrc2,epsfig]{article}

\title{Heavy light mesons at a larger lattice spacing
}
\author{Joachim Hein%
\address{Department of Physics \& Astronomy, Glasgow G12 8QQ, UKQCD Collaboration.} 
for the GLOK Collaboration%
\thanks{In Collaboration with A.~Ali~Khan, T.~Bhattacharya, S.~Collins,
C.T.H.~Davies, R.~Gupta, C.~Morningstar, J.~Shigemitsu, J.~Sloan}
}

\newcommand{\DD}{{\rm D}}

\newcommand{\fin}{\hspace{0.6cm}}
\newcommand{\imag}{{\rm i\hspace{0.13ex}}}

\newcommand{\bgeq}{\begin{equation}}
\newcommand{\bgeqa}{\begin{eqnarray}}
\newcommand{\edeq}{\end{equation}}
\newcommand{\edeqa}{\end{eqnarray}}

\newcommand{\ainv}{a^{-1}}

\begin{document}

\begin{abstract}
We present results on the spectrum of $B$ and $D$ mesons including
radial excitations and discuss the pseudoscalar decay constant.
The results are obtained at $\beta=5.7$ in the 
quenched approximation using NRQCD for the heavy quark. To study
scaling violations we also compare to results obtained at $\beta=6.0$.
\end{abstract}

\maketitle

\noindent\hspace*{-1.9mm}
\raisebox{6cm}[0ex][0ex]{
{\normalsize
\parbox{4cm}{
\textbf{\textsf{GUTPA/97/10/1}}\\
\textbf{\textsf{hep-lat/9710097}}}}
}\vspace*{-5ex}
\section{INTRODUCTION}
We want to detail our new and as yet preliminary
simulation  results on the spectrum and the renormalised decay
constant of heavy light mesons. They were obtained on coarser lattices
compared to previous analyses, see \cite{arifa} for a review. This
allows us to study scaling violations. A second motivation is the
possibility of reaching 
the charm region with NRQCD for $D$ mesons, which have a different power
counting than $J/\psi$. 

For the simulations we use 
278 quenched $12^3\times 24$ configurations at $\beta=5.7$
generously provided by the UKQCD-Collaboration. 
To simulate the light quarks on the lattice 
we use the tadpole improved clover action with
$\kappa_c=0.1434$ from \cite{hugh} and
$\kappa_s|_{K}=0.1398$ from \cite{Prowland}.
The Hamiltonian of the heavy quarks is expanded around 
its non relativistic limit 
with tadpole improved gauge links and fields:
\bgeq \label{hamilton}
H = H_0 + \delta H\,,
\edeq
were:
\bgeqa
H_0 &:=& - \frac{\DD^2}{2M_0}\,,\\[1ex]
\delta H &:=& -\frac{g}{2M_0}{\vec \sigma\, \vec B} +
\frac{\imag g}{8M_0^2}(\vec \DD\,\vec E - \vec E\, \vec \DD)\nonumber\\
&& - \frac{g}{8M_0^2}\vec \sigma(\vec \DD \times \vec E - \vec
E\times\vec \DD)\nonumber \\
&&-\frac{(\DD^2)^2}{8M_0^3}+a^2 \frac{\DD^{(4)}}{24M_0}
-a\frac{(\DD^2)^2}{16nM_0^2}\,.\label{deltah}
\edeqa
This expansion is correct up to ${\cal O}(M_0^{-2})$ with the last two
terms being corrections for finite lattice spacing.
In the quenched approximation heavy or light quark physics deliver
different lattice spacings:
from $m_\rho$ one gets 
$\ainv=1.103(11)(50)$~GeV \cite{hugh} and from the $\Upsilon$-spectrum
1.4(1)~GeV \cite{scaling}.

\section{MESON SPECTROSCOPY}
So far we have analysed correlators with the $^1S_0$ and $^3S_1$ quantum
numbers and use two different smearing functions for each value of
$M_0$. When extracting masses we fit with two exponentials in
case of the respective ground state and three in case of radial excitations.
The position of the $c$ and the $b$ quark in the $M_0$ range is
fixed from the requirement that the $D$ and $B$ masses must coincide
with 
their physical values in GeV \cite{pdg}. 
To correct the meson mass for the omitted $M_0$ in (\ref{hamilton})
we use perturbative shifts \cite{morningstar}, which are
in very good agreement with the non perturbative shifts from $J/\psi$
and $\Upsilon$ \cite{charm,scaling}.

\begin{figure}[tb]
\epsfig{file=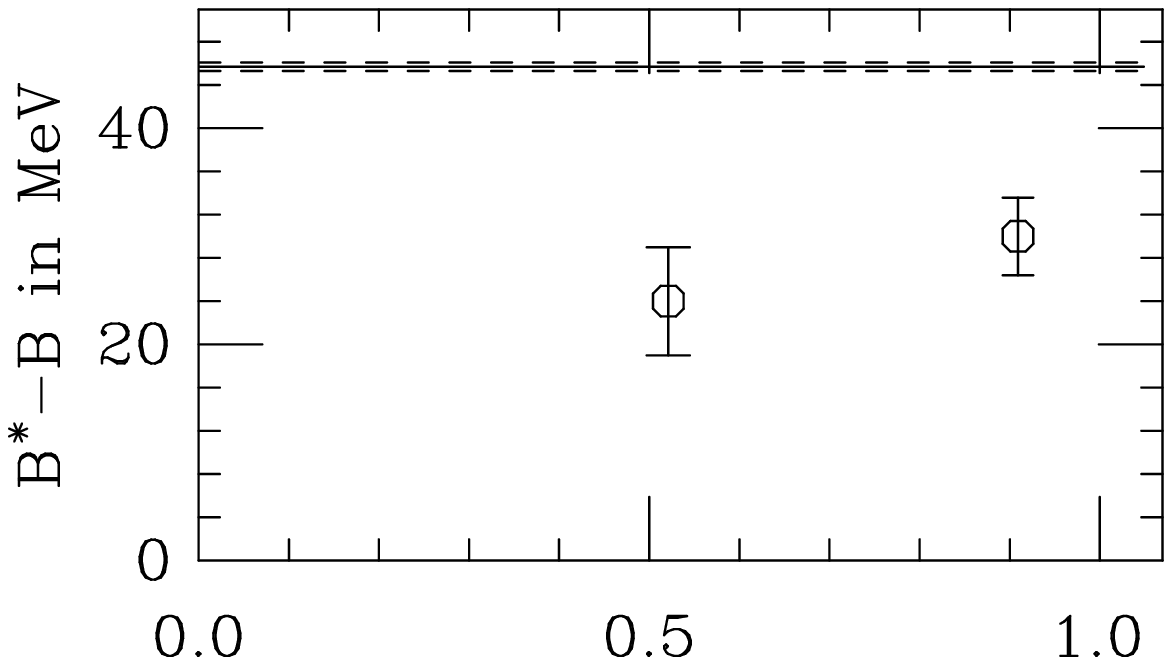,width=4.65cm,bbllx=35pt,bburx=390pt,bblly=70pt,bbury=265pt}
\epsfig{file=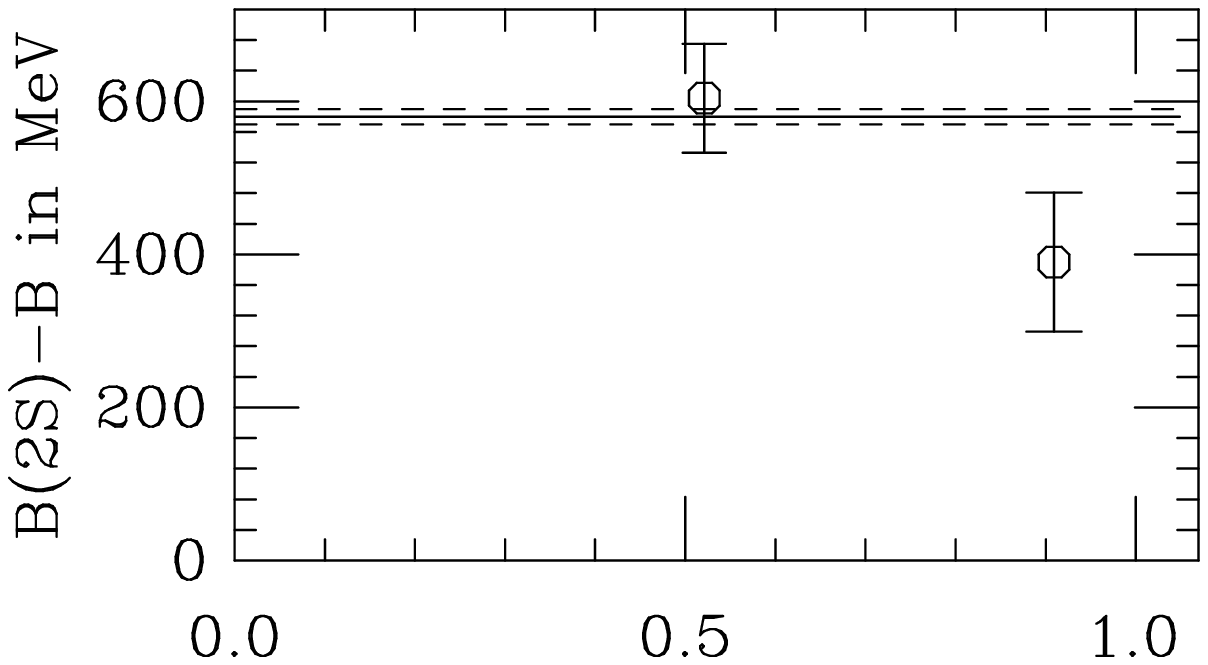, width=4.65cm,bbllx=35pt,bburx=390pt,bblly=70pt,bbury=265pt}
\epsfig{file=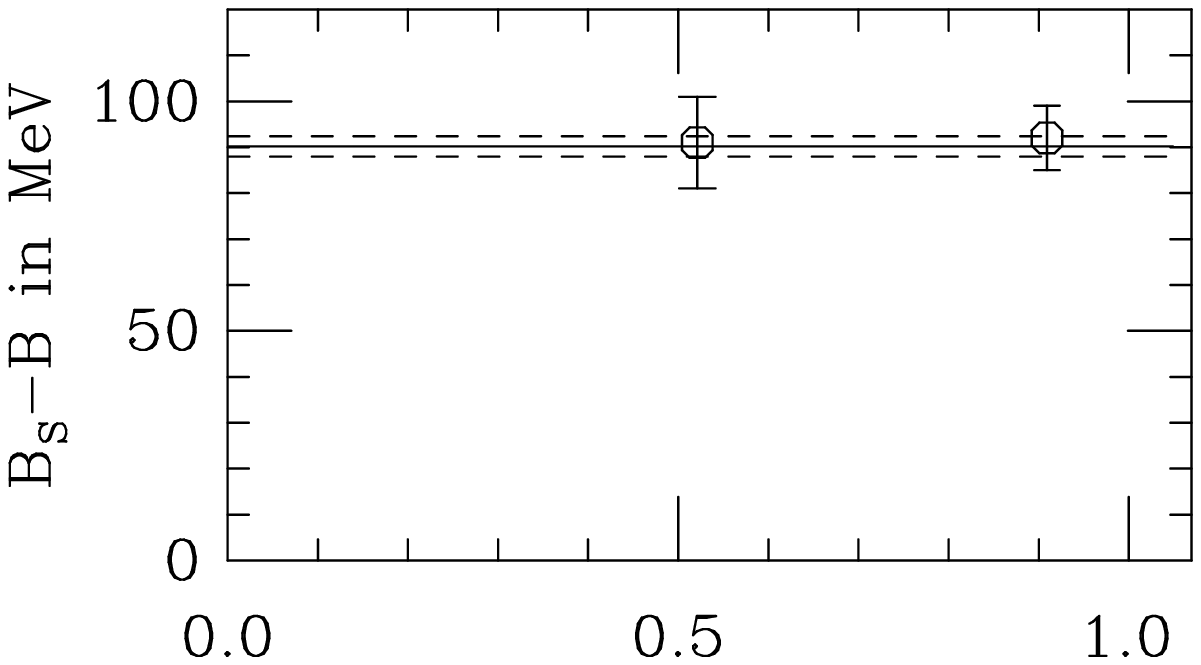, width=4.65cm,bbllx=35pt,bburx=390pt,bblly=70pt,bbury=265pt}
\epsfig{file=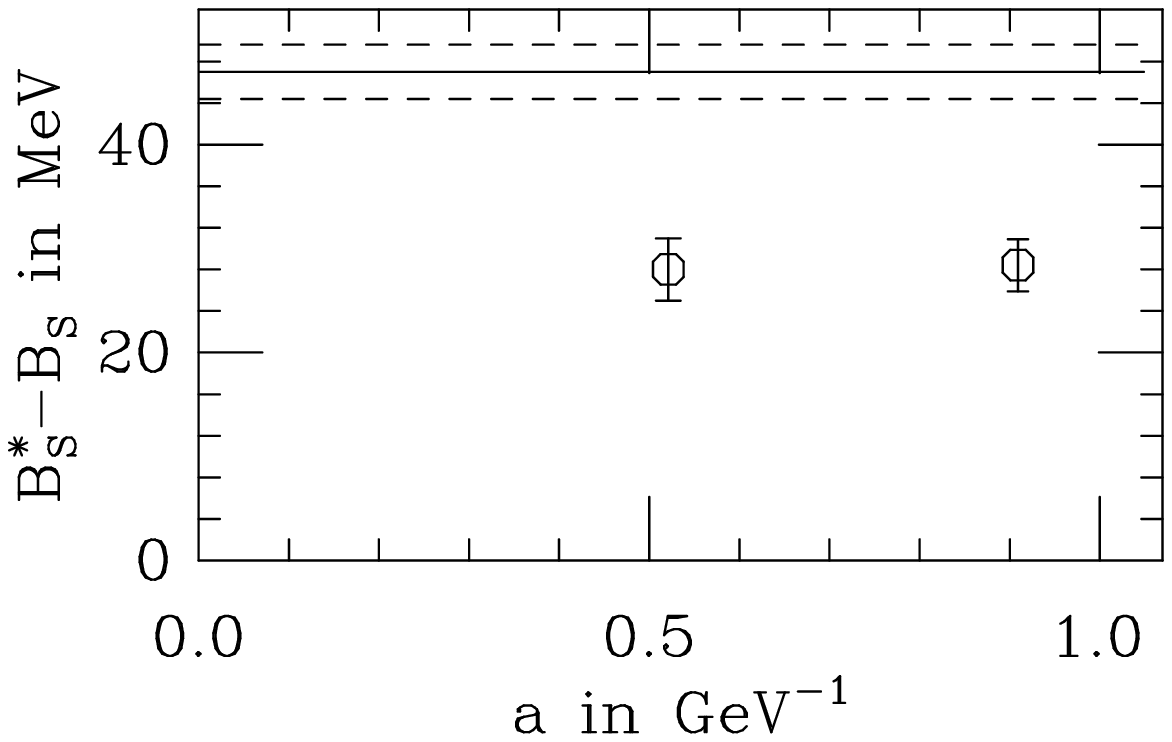, width=4.65cm,bbllx=35pt,bburx=390pt,bblly=50pt,bbury=265pt}
\caption{\label{bspecfig}Scaling of the $B$ spectrum with $a$.}
\end{figure}

In fig.~\ref{bspecfig} we give our results for the $B$
spectrum together with preliminary results at $\beta=6.0$ with the
same action.
For the comparison we set the scale from 
$m_\rho$, since the uncertainty in how to determine $a$
cancels out between both lattice results. 
The error bars represent the statistical uncertainties and
those arising in the determination of the bare $b$ quark mass.
The horizontal lines give the experimental values, quoted from
\cite{pdg}, with exception of the radially excited $B(2S)$, where we
quote a preliminary {\it DELPHI\/} result \cite{feindt}.
Comparing the results, no significant scaling violation can be
observed within the error bars of $\approx
10\%$ in case of the hyperfine and $B_s-B$ splitting.

\begin{figure}[tb]
\epsfig{file=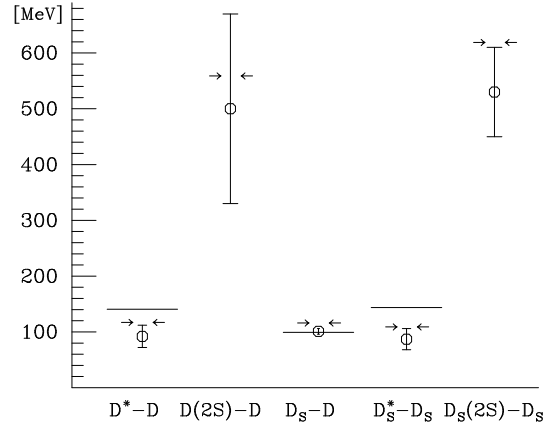,width=7.0cm,bbllx=45pt,bburx=505pt,bblly=104pt,bbury=470pt}
\caption{\label{dspecfig}
Splittings in the $D$ meson spectrum. The circles give the result for 
$a^{-1}=1.10$~GeV and the arrows for 1.25~GeV. The experimental values
are given by horizontal lines.}
\end{figure}

Our result for the $D$ spectrum is given in
fig.~\ref{dspecfig}. 
The arrows indicate the shift of the central values when taking the
average of the above quoted $a$-values instead of $a(m_\rho)$, the
other symbols are similar to fig.~\ref{bspecfig}.
When comparing to the experimental values, the 
splittings between the strange and non strange mesons 
are in very good agreement and
also the result for the radial excited $B(2S)$ agrees within its fairly
large error bar. However all hyperfine splittings are much too low when
using $a$ from $m_\rho$. When using a larger value of $\ainv$, for
example from $\Upsilon$, the hyperfine increases
but the agreement of the $B_s - B$ and $D_s-D$ is lost. The cause of
this trouble might be a systematic error of the quenched
approximation, a quite large 1-loop correction to the coefficient in
front of the $\vec \sigma\,\vec B$ term or neglected
${\cal O}(M^{-3})$ terms in eq.~(\ref{deltah}). This
question definitely requires further investigation.

It is interesting to have a look at the ratio of the hyperfine
splittings. We find: 
\bgeq \label{hyperratio}
\frac{D^* -D}{B^*-B} = 3.1(7)\,,\fin\frac{D^*_s -D_s}{B^*_s-B_s} = 3.1(7)\,,
\edeq
which has to be compared to an experimental value of 2.95(2) resp.\
3.06(17). The error in (\ref{hyperratio}) is dominated by the 
uncertainty of the perturbative mass shifts in the determination of
the bare $c$ quark mass. However this might be too conservative and
the agreement of the central values itself
is quite impressive.

\section{DECAY CONSTANT}
In the effective theory one has to measure the matrix elements of the
currents
\bgeqa
J_A^{(0)} &=& \bar q \gamma_5 \gamma_0 Q\,,\nonumber\\
J_A^{(1)} &=& -\frac{1}{2M_0}\bar q \gamma_5 \gamma_0 (\vec\gamma\vec D)
Q\,,\nonumber\\
J_A^{(2)} &=& \frac{1}{2M_0} (\vec D \bar q \cdot \vec\gamma)\gamma_5 \gamma_0 Q\,,
\edeqa
in order to calculate the pseudoscalar decay constant up to 
${\cal O}(M^{-1}_0)$. On the lattice for zero momentum we have
$J_A^{(1)}=J_A^{(2)}$.
Their renormalisation and
matching to full QCD have recently been calculated to 1-loop 
order \cite{JunkoT,Junko}.
However as yet the matching scale $q^*$ has not been
calculated and we will give results for $aq^* = 1$ and $\pi$. 
We will also apply the ${\cal O}(a \alpha_s)$ improvement term as
introduced in the references. 
For the $\log (aM)$ terms in the coefficients of 
$J_A^{(0)}$ and $J_A^{(1)}$ we insert $\frac{\alpha_s}{\pi}\log(aM_0)$, which
will give the physical decay constant for $B$ and $D$ mesons correctly.

\begin{figure}[tb]
\epsfig{file=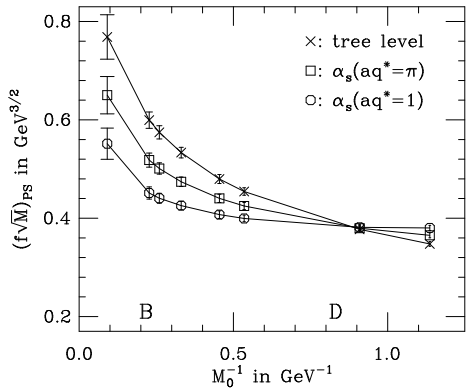,width=6.1cm,bbllx=58pt,bburx=508pt,bblly=45pt,bbury=423pt}
\caption{\label{fren57}
Pseudoscalar decay constant. The connecting lines are to
guide the eye. The position of the 
$B$ and the $D$ meson are indicated. Statistical errors only are shown.}
\end{figure}

\begin{figure}[tb]
\epsfig{file=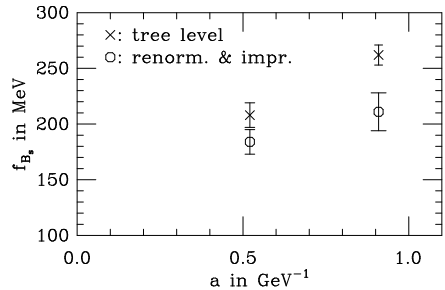,width=5.7cm,bbllx=45pt,bburx=466pt,bblly=52pt,bbury=329pt}
\caption{\label{fscalefig}Scaling of $f_{B_s}$. The error bars contain
statistical errors and the uncertainty of $q^*$.}
\end{figure}

So far we have analysed the currents for $\kappa = 0.1400$ which is
close to $\kappa_s|_K$. From our experience at smaller $a$ we
expect the effect of this mismatch to be $1\%$  on $f_{\rm PS}$,
which is negligible compared to the other errors. 
Our result is given in fig.~\ref{fren57}. 
In the $B$ region
$f_{\rm PS}$ is reduced quite largely by 15-25\%  
due to renormalisation and improvement. We obtain
\bgeq
f_{B_s} = 211(7)(15)(37){\rm \ MeV}\,.
\edeq
In parenthesis we give the statistical uncertainty, the uncertainty of
$q^*$ and the shift of the central value for $a^{-1}=1.25$~GeV instead
of 1.10~GeV.
In fig.~\ref{fscalefig} we discuss the scaling of this quantity.
Since the correct $q^*$ values need not to be equal 
at $\beta=5.7$ and 6.0,
we average over $aq^*=1$ and $\pi$. The difference is included 
into the error bars.

At tree level we observe a scaling violation of $\approx 30\%$
which is reduced in case of the renormalised and improved
result, so that they agree within error bars. Note that the result at
larger $a$ is more sensitive to $q^*$.

\section{CONCLUSION}
We observe no significant scaling violations for the
$B$ spectrum and the renormalised $f_{B_s}$. Our results on the $D$
spectrum look encouraging.

A personal fellowship by the European commission under ERB FMB ICT 961729,
and financial support of the collab.\ by
NATO under CRG 941259 and by the US DOE.\ is gratefully acknowledged.
The simulations were performed at NERSC.

\end{document}